\documentclass[aps,prstab,superscriptaddress,longbibliography, twocolumn,amsmath,amssymb]{revtex4-2}
\usepackage{graphicx}
\usepackage{dcolumn}
\usepackage{graphicx}
\usepackage{dcolumn}
\usepackage{hyperref}

\hypersetup{
    colorlinks=true,
    linkcolor=blue,
    citecolor=blue,
    filecolor=magenta,
    urlcolor=blue
}
\newcommand{\Figref}[1]{Fig.\ref{#1}}
\newcommand{\Tabref}[1]{Table~\ref{#1}}
\newcommand{\bR}{\mathbf{R}}
\newcommand{\citenamefont}{}
\newcommand{\bibnamefont}{}

\begin{document}

\title{A COMPACT, EASY-TO-USE LEAK DETECTOR FOR VACUUM SYSTEMS BASED ON THE MQ-8 SENSOR}

\author{S.M. Kovalov}
\altaffiliation[Corresponding author ]{\\Email address: covalov.sergiy@gmail.com (S.M. Kovalov)}

\author{I.V. Beznosenko}
\author{A.V.~Vasyliev}
\author{G.V.~Sotnikov}

\address{National Science Center Kharkiv Institute of Physics and Technology \\
Institute of Plasma Electronics and New Acceleration Methods \\
1, Akademichna St., Kharkiv, 61108, Ukraine}

\date{\today}

\begin{abstract}
This paper presents the development of a compact leak detector for vacuum systems that operates using hydrogen as a tracer gas, detected by a semiconductor MQ-8 sensor. The sensor is connected to an Arduino microcontroller, enabling digital signal processing and real-time visualization through the Processing software environment. The designed device is capable of detecting even small hydrogen leaks caused by imperfect sealing of vacuum connections. It is characterized by simplicity, low manufacturing cost, and suitability for laboratory use as an alternative to helium leak detectors. Experimental tests confirmed the effectiveness and stability of the developed leak detector during long-term measurements.

\end{abstract}

\maketitle

\section{INTRODUCTION}
In experiments involving electron acceleration within vacuum chambers, particularly in dielectric laser accelerator (DLA) systems, a stable high vacuum is essential to minimize scattering and ensure stable propagation of the electron beam. For conducting physical experiments on DLA, the computer simulation of which is described in \cite{Vasyliev2024,Beznosenko2023,Beznosenko2024}, the operating pressure in the vacuum chamber must be maintained at the level of $10^{-4}$--$10^{-5}$ Pa. At such low pressures, the mean free path of gas molecules greatly exceeds the physical dimensions of the vacuum chamber, allowing the electron beam to pass through with negligible energy loss or defocusing. The mean free path $\ell$ is determined by an expression derived from the kinetic theory of gases \cite{Landau1981,TecScience}:

\begin{equation}\label{eq:01}
\ell = \frac{k_{B}T}{\sqrt{2}d^{2}p\pi},
\end{equation}

where $k_{B}$ is the Boltzmann constant, $T$ is the gas temperature in the chamber, $d$ is the effective diameter of an air molecule, $p$ is the pressure.

In our experiments, the operating pressure in the chamber is on average $p = 5 \times 10^{-5}$ Pa at a temperature of $T=293$ K, which corresponds to high-vacuum conditions. For air, which consists of 99 \% nitrogen and oxygen molecules, the effective molecular diameter is taken as $d = 3.6 \times 10^{-10}$ m \cite{Kretzschmar2022}. Substituting these values into equation (\ref{eq:01}), we obtain the mean free path of the gas molecules of approximately 137 m, which is many times greater than the characteristic dimensions of the used vacuum chamber (0.5 m). This indicates an extremely low probability of collisions between electrons and residual gas molecules. Consequently, the electron beam can propagate along the entire length of the chamber without significant energy losses or trajectory deviations, ensuring the stability and reproducibility of the experiment.

During one of the experiments with the electron gun, a decrease in the VIT-2 vacuum gauge reading to 1.3 mV was recorded, which, according to its calibration curve, corresponds to a pressure of about $1.5\times10^{-1}$ mmHg ($\approx$20 Pa). At this pressure, in a chamber with a volume of 200 L, air electric breakdown processes begin to occur, making it impossible to conduct experiments with the electron gun. This prompted a diagnostic check of the vacuum system to identify the location of the leak.

\section{ANALYSIS OF EXISTING METHODS AND DEVELOPMENT PREREQUISITES}
In modern vacuum technology, there is a wide range of methods for leak detection. The most common and highly accurate solution is helium mass spectrometer leak detectors, with sensitivities reaching $10^{-11}$--$10^{-12}$ m$^3\cdot$Pa/s. Their operating principle is based on the detection of helium atoms that penetrate the vacuum system through microdefects and are measured via the ion current in the mass spectrometer. Such systems are used for the certification and diagnostics of sealed setups in the electronics, cryogenic, and nuclear industries \cite{PriboryOnline,Pfeiffer2013}. However, the practical use of helium leak detectors in laboratory conditions, especially under limited resources, is associated with a number of significant challenges. First, these instruments are expensive: the price of typical models on the Ukrainian market is at least around USD 2,500 \cite{PriboryOnline}. Second, their operation requires constant access to a source of high-purity helium (99.999\%) \cite{Pfeiffer2013,INFICON2023}. Additionally, regular calibration and maintenance of vacuum pumps are necessary, which further increases operational costs.

Moreover, mass spectrometer-based systems are bulky and heavy (15--25 kg) \cite{PriboryOnline}, making them inconvenient for rapid use in laboratories with limited space. Their start-up and stabilization to operational mode take several minutes, and monitoring is primarily stationary. In the case of a sudden vacuum loss or the need for rapid diagnostics, such systems prove inefficient, as their use consumes significant time and resources.

In practical vacuum research, classical leak detection methods, such as blowing with volatile liquids (alcohol, freon) or applying soapy solutions, are used to localize leaks at moderate pressures. However, their effectiveness sharply decreases in the case of significant leaks, since the system pressure becomes too high to detect small changes in vacuum readings. Under these conditions, the evaporation of liquid indicators occurs too quickly, and the sensitivity of standard thermocouple vacuum gauges (VIT-2, VIT-1, VIT-3) is insufficient to register such changes (in our case, at a pressure of about 20 Pa, the ionization-thermocouple gauge VIT-2 did not detect any signal fluctuations).

Furthermore, the use of soapy solutions is often limited by the design features of the equipment. If the setup has a complex geometry or double walls (inner and outer, as in the considered vacuum chamber), the space between them becomes inaccessible for visual inspection. This makes it impossible to use foaming methods to determine the leak location.

Thus, in our case, traditional methods for leak diagnostics are ineffective in the presence of significant leaks and a complex vacuum chamber design. The absence of a helium mass spectrometer leak detector and the impracticality of acquiring one quickly in a laboratory setting may necessitate the search for an alternative approach capable of providing sufficient sensitivity at low cost and minimal preparation time.

Below, we describe the development of a compact leak detector technology that can be quickly fabricated in laboratory conditions without significant material expenses.

\section{HYDROGEN AS A TRACER GAS AND MQ-8 SENSOR AS DETECTOR}
An analysis of existing approaches for detecting leaks in vacuum chambers has shown that using hydrogen as a tracer gas represents an optimal compromise between sensitivity, cost, and ease of implementation. This method has been successfully applied in a number of industrial and research setups \cite{Fakra2020}, confirming its reliability and reproducibility.

Hydrogen has a minimal molecular diameter (about 2.9 Å) \cite{Halter2020} and high diffusivity, allowing it to penetrate even the tiniest defects in vacuum joints. Furthermore, the gas can be easily generated using a laboratory electrolyzer, eliminating the need for expensive cylinders and auxiliary equipment. Thus, a hydrogen-based method provides a combination of high efficiency, compactness, and low cost, while the supply of hydrogen in small flows ($\sim$10 L/h) ensures safe operation and eliminates the risk of explosion, making the method particularly promising for laboratory conditions.

For hydrogen detection, a semiconductor sensor MQ-8 was selected. Its operation is based on the change in resistance of the tin dioxide (SnO$_2$) sensitive layer upon interaction with reducing gases \cite{Fakra2020}. This sensor is low-cost, easy to interface, and sufficiently sensitive (in the range of 100 -- 10,000 ppm \cite{Fakra2020}) for detecting leaks in laboratory settings. When combined with an Arduino Uno microcontroller, it enables continuous monitoring of gas concentration and real-time data visualization using the Processing software environment.

\section{OPERATING PRINCIPLE OF SEMICONDUCTOR SENSORS (MQ FAMILY)}
MQ-type sensors implement a chemiresistive principle. The sensitive element consists of a semiconductor film, typically based on SnO$_2$ with catalytic additives \cite{Fakra2020}, which is heated by an integrated heater to several hundred $^\circ$C. In the presence of reducing gases (H$_2$, etc.), molecules adsorb onto the surface, altering the charge carrier density in the near-surface layers. This leads to a change in the resistance $R_S$ of the sensitive element. Thus, the gas concentration is converted into an electrical signal.

According to the manufacturer's technical documentation (MQ-8 Datasheet) \cite{MQ8Datasheet}, the sensor has the following characteristics. The operating temperature of the sensitive element is maintained by the built-in heater and should be 300 $^\circ$C. A warm-up period of at least 60 seconds is required after power-on to stabilize the sensor's characteristics. The sensitivity curve is nonlinear and is described by a power-law dependence $\frac{R_{s}}{R_{0}}$ on hydrogen concentration in the range of 100 ppm to 10,000 ppm \cite{Fakra2020}. The sensor demonstrates high selectivity to hydrogen, but exhibits cross-sensitivity to carbon monoxide and methane, as indicated by the standard curves in the datasheet.

An important operational factor is the drift of the sensor's resistance over time due to aging of the sensitive layer and exposure to interfering gases, which necessitates periodic recalibration. In addition, the readings are influenced by ambient temperature and humidity. Therefore, the manufacturer recommends conducting measurements at relative humidity not exceeding 85\% and temperatures between 20 and 40 $^\circ$C.

Unlike mass-spectrometric helium leak detectors that operate on the physical principle of detecting helium ions, the MQ-8 sensor is based on changes in the electrical conductivity of a semiconducting SnO$_2$ layer upon interaction with hydrogen molecules. This approach provides a simple design, low cost, and the possibility of implementing a compact digital device with sufficient sensitivity for laboratory applications.

\section{ELECTRICAL MODEL AND CONNECTION SCHEME OF THE MQ-8 SENSOR}
The sensor output is formed by a voltage divider $V_{out}$ across $R_S$ (the sensor resistance) and an external resistor $R_L$ ("load resistor"), connected to ground:

\begin{equation}\label{eq:02}
V_{out} = V_{CC} \cdot \frac{R_{L}}{R_{L} + R_{s}},
\end{equation}

which describes a voltage divider, where $V_{CC}$ is the supply voltage (typically 5 V), $R_L$ is the external load resistor, and $R_S$ is the resistance of the sensor's sensitive element, which depends on the hydrogen concentration. Under normal conditions, at low hydrogen concentration $R_S$ is high, so the voltage drop across the load resistor is small, and the output voltage $V_{out}$ remains low. As the hydrogen concentration increases, $R_S$ decreases, leading to a rise in the output voltage (\ref{eq:02}). Thus, changes in the chemical composition of the air near the sensor are directly converted into an electrical signal suitable for measurement. The MQ-8 sensor output can be connected to an analog input of an Arduino Uno microcontroller (e.g., pin A0). The microcontroller contains a built-in 10-bit analog-to-digital converter (ADC), which allows converting voltages from 0 to 5 V into digital codes ranging from 0 to 1023. This means that small changes in the sensor output voltage can be detected with a resolution of approximately 4.9 mV \cite{Fakra2020}.

For more accurate detection of weak signals, the internal reference voltage of the Arduino (1.1 V) can be used, which increases the ADC sensitivity almost fivefold. In this case, the value of the load resistor $R_L$, must be adjusted so that the sensor output voltage does not exceed 1.1 V even at maximum hydrogen concentration.

The operation of the entire system can be described as follows. The built-in heater of the MQ-8 sensor maintains a high temperature of the sensitive layer, enabling adsorption and desorption of gases. When the hydrogen concentration changes, the layer resistance changes, and according to the voltage divider formula, the output voltage varies. This voltage is converted by the Arduino ADC into a digital code, which can then be processed in software: averaged, filtered, and analyzed to detect threshold exceedances.

Thus, the combination of the sensor's analog model and digital processing allows detection of both large leaks (hydrogen concentrations above 1,000--2,000 ppm) and small leaks, where hydrogen changes are on the order of 50--100 ppm relative to the background level. Visualization in the Processing environment \cite{Processing} further facilitates operation by displaying real-time concentration changes as graphs.

\section{METHOD FOR LEAK DETECTION AND LOCALIZATION}
\begin{figure}[!bh]
  \centering
  \includegraphics[width=0.48\textwidth]{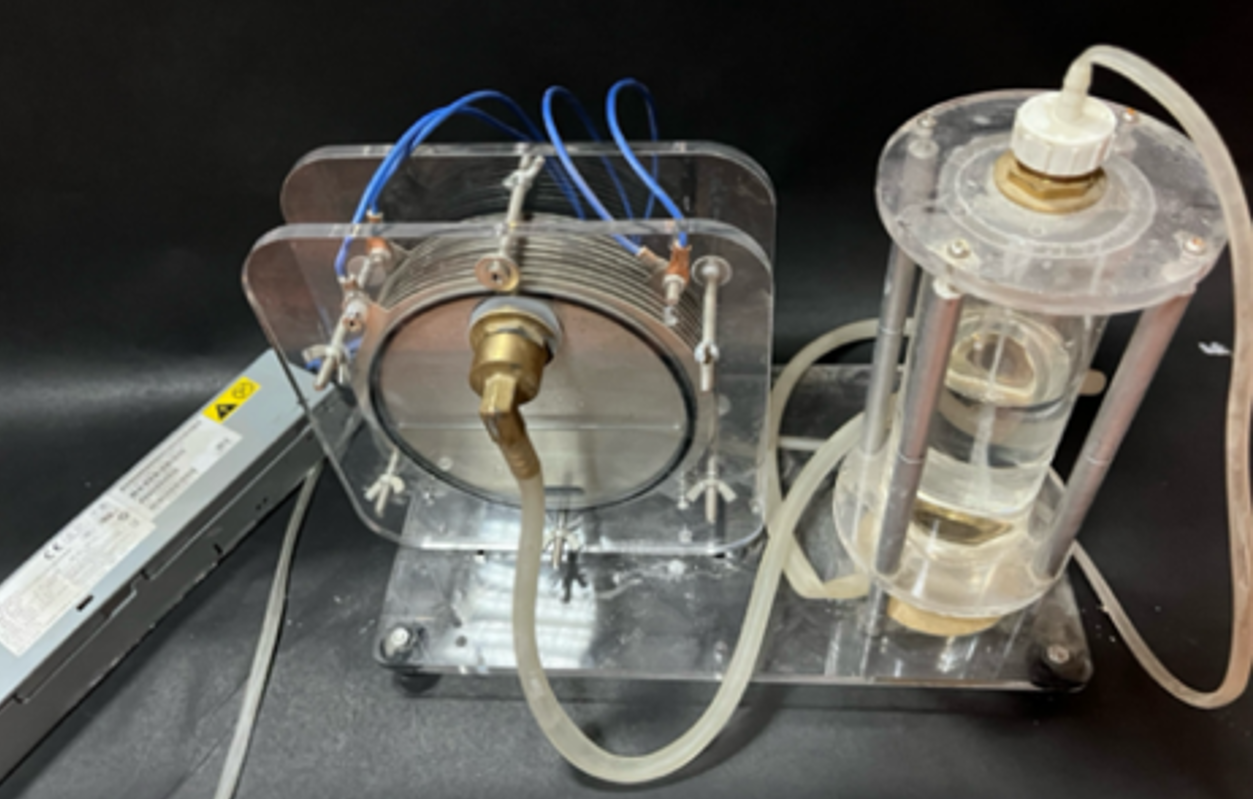}
  \caption{Laboratory hydrogen electrolyzer used in the experiment.}
  \label{Fig:01}
\end{figure}

To localize leaks in the vacuum system, a gas-blowing method was implemented using hydrogen as a tracer gas. The hydrogen source was a laboratory electrolyzer (\Figref{Fig:01}), consisting of a unit with plate electrodes immersed in an aqueous electrolyte and a separate gas collection flask. When a direct current was applied to the electrodes, electrolysis of water occurred, releasing gaseous hydrogen at the cathode. The electrolyzer design allowed continuous operation and enabled adjustment of the gas generation rate by varying the current. The gas outlet was connected to a fitting with a flexible hose used to blow the external surface of the chamber. To prevent overpressure and enhance safety, a water seal (acting as a check valve) was employed. Hydrogen was supplied to the system as a slow and steady flow, which allowed controlled blowing and minimized background fluctuations.

The gas was directed onto the outer surface of the vacuum chamber in the area of the suspected leak. At the same time, the forevacuum pump created a pressure drop inside the system, ensuring air intake through the defect. If a leak was present in the blown area, part of the hydrogen penetrated into the chamber and was then carried by the air flow toward the forevacuum pump outlet.

\begin{figure}[!bh]
  \centering
  \includegraphics[width=0.48\textwidth]{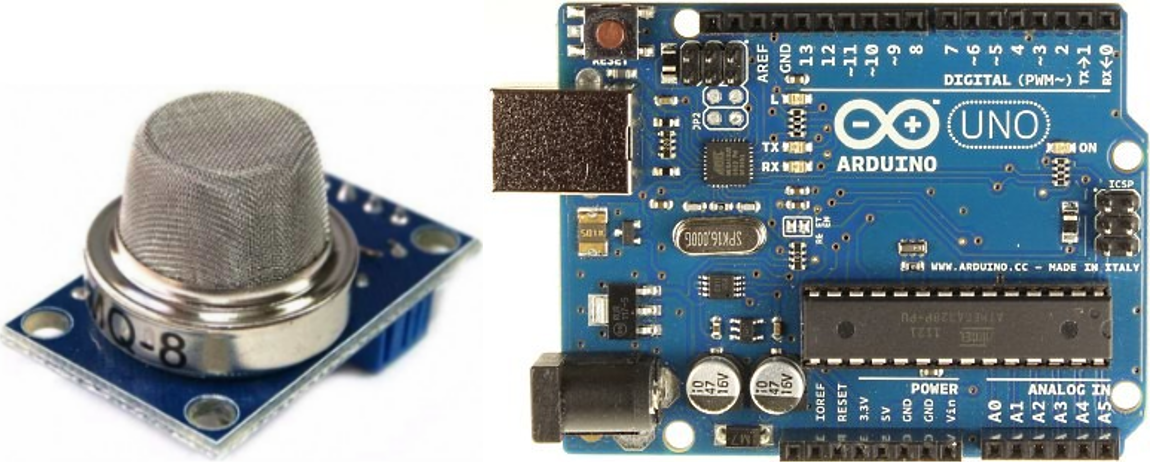}
  \caption{MQ-8 sensor and Arduino microcontroller.}
  \label{Fig:02}
\end{figure}

\begin{figure}[!bh]
  \centering
  \includegraphics[width=0.48\textwidth]{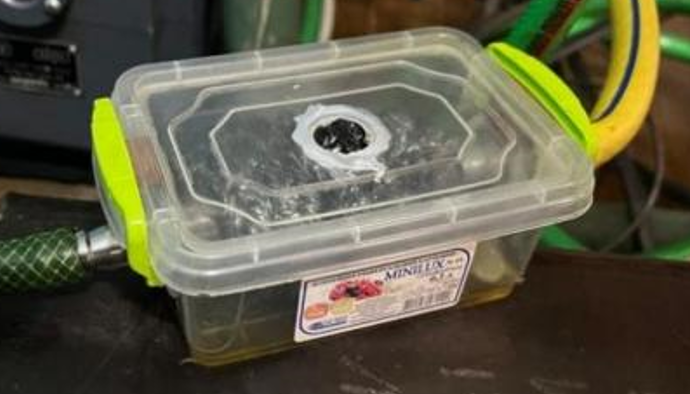}
  \caption{Additional reservoir installed at the outlet of the forevacuum pump.}
  \label{Fig:03}
\end{figure}

\begin{figure}[!bh]
  \centering
  \includegraphics[width=0.48\textwidth]{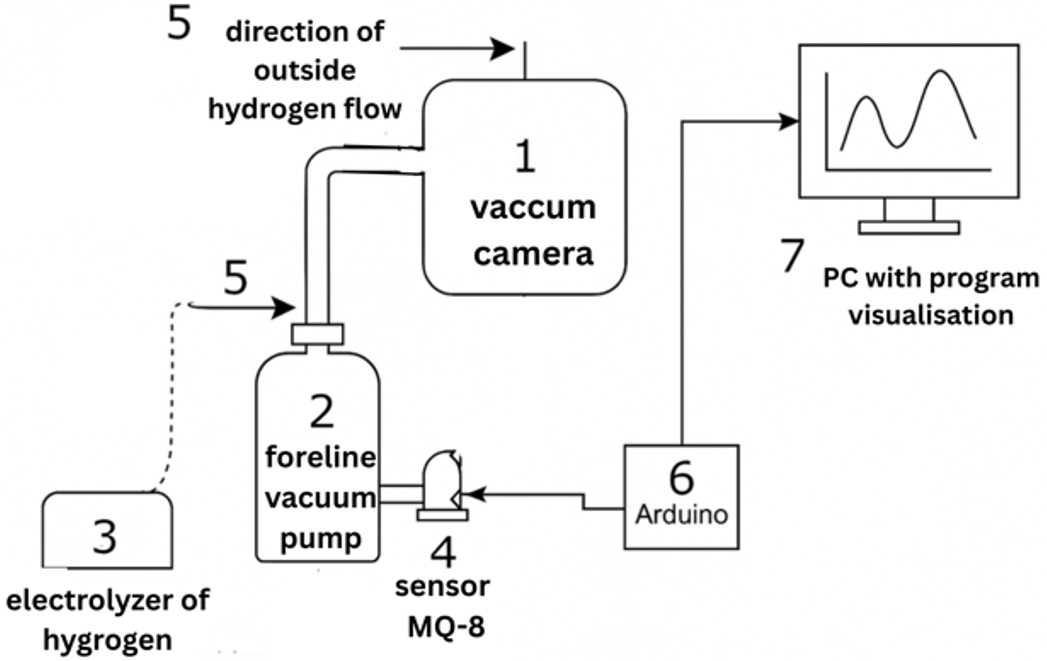}
  \caption{Diagram for leak detection in the chamber: 1 is vacuum chamber; 2 is forevacuum pump; 3 is hydrogen electrolyzer; 4 is MQ-8 sensor; 5 is direction of hydrogen flow; 6 is Arduino (digitization and data transfer); 7 is personal computer with visualization software.}
  \label{Fig:04}
\end{figure}

At the outlet line of the forevacuum pump, an MQ-8 sensor (\Figref{Fig:02}) was installed, operating in the continuous hydrogen concentration monitoring mode. To increase sensitivity, an additional reservoir (\Figref{Fig:03}) was used, in which the gas flow rate decreased and partial accumulation of hydrogen occurred. The sensor was placed in the upper part of this reservoir, which ensured more reliable detection of gas concentration. Its electrical signal was read by an Arduino Uno microcontroller \cite{Arduino} (\Figref{Fig:02}) at a sampling frequency of about 10 Hz, after which the data in digital form were transmitted to a personal computer via a USB interface (\Figref{Fig:04}).

To improve the informativeness of measurements, digital filtering algorithms implemented on the Arduino microcontroller were used. The main data processing tool was the Exponential Moving Average (EMA), which allowed smoothing out rapid noise fluctuations of the sensor output signal. The algorithm can be expressed as a recurrent equation:

\begin{equation}\label{eq:03}
S_{t} = \alpha \cdot X_{t} + (1 - \alpha) \cdot S_{t - 1},
\end{equation}

where $S_t$ is the smoothed signal value at the current time, $X_t$ is the current measurement from the sensor, and $\alpha$ is the smoothing coefficient ($0 < \alpha < 1$).

To eliminate false spikes, a jump limiter was applied. When a sharp change in readings exceeded the preset threshold $\Delta_{\max}$ the value was corrected according to the formula:

\begin{equation}\label{eq:04}
X_{t}' = X_{t - 1} + \text{sign}\left( X_{t} - X_{t - 1} \right) \cdot \Delta_{\max},
\end{equation}

which made it possible to minimize the effect of random impulsive noise.

The threshold for sensor triggering was set dynamically and defined as the sum of the baseline level and a value proportional to the current standard deviation of the signal:

\begin{equation}\label{eq:05}
T = B + k \cdot \sigma,
\end{equation}

where $T$ is the threshold, $B$ is the baseline signal value, $\sigma$ is the standard deviation over a fixed time interval, and $k$ is the coefficient determining system sensitivity. This approach allowed adaptive consideration of background variations and increased the reliability of detecting low hydrogen concentrations.

Data visualization was performed in the Processing environment (\Figref{Fig:05}) \cite{Processing}. On the computer screen, a real-time graph of the sensor signal changes was displayed. When hydrogen entered the system, a characteristic spike significantly exceeding the level of background fluctuations was registered. This allowed the operator to promptly record the moment of leakage and adjust the position of the blowing point. A software-controlled sequence of blowing different sections of the chamber surface was implemented. The appearance of a sharp signal spike during hydrogen supply to a specific area made it possible to localize the defect with high accuracy and confirm the correctness of the applied method.

\section{EXPERIMENTAL RESULTS}
During the tests, characteristic changes in the sensor output signal were recorded when blowing specific areas of the chamber. The reproducibility of the response during hydrogen supply confirmed the presence of a leak.

\begin{figure}[!bh]
  \centering
  \includegraphics[width=0.78\textwidth]{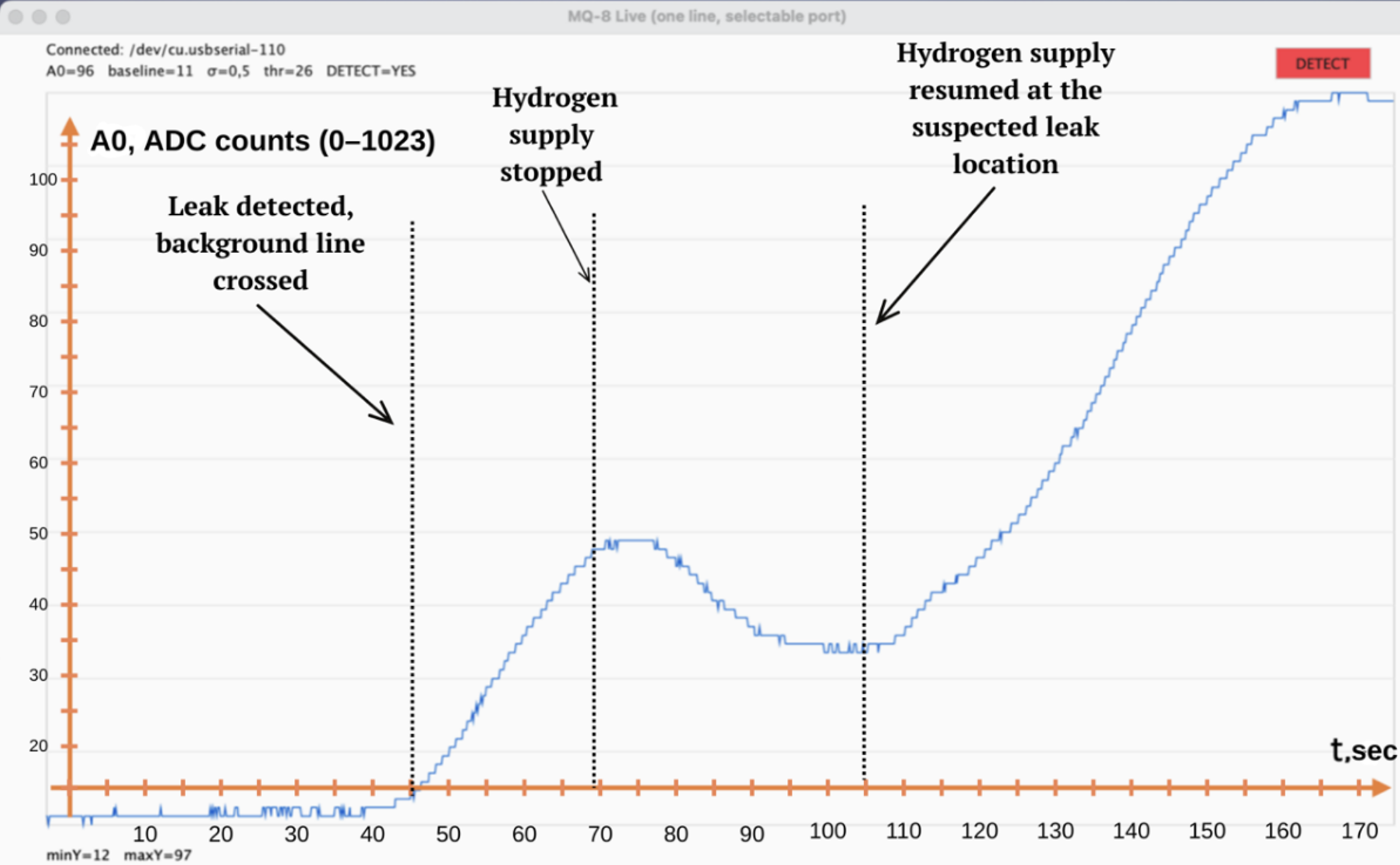}
  \caption{Dependence of the Arduino analog output signal (ADC, from 0 to 1023) on time.}
  \label{Fig:05}
\end{figure}

A typical example of real-time observation of the signal during leak detection is shown in \Figref{Fig:05}. Along the t-axis (horizontal), time was plotted, and along the A0-axis (vertical), the analog value of the sensor output signal was plotted. Before the start of measurements, the sensor was software-calibrated to establish the baseline signal value corresponding to normal conditions without hydrogen (in our case, this value was 15). When this level was exceeded (the curve going above the baseline), the program registered the event "DETECT = YES" and marked the leak occurrence.

Thus, the change in hydrogen concentration was visualized as a characteristic spike on the graph, allowing quick identification of a vacuum chamber defect at the sensor's position and evaluation of the sensor's sensitivity. At the 45th second, the baseline level of the analog signal (15 counts) was crossed, indicating the presence of a leak. The graph showed a gradual increase in the signal level for about 30 seconds. After a temporary interruption of hydrogen supply at the 69th second, the analog signal dropped from 48 to 34 counts, and when hydrogen was supplied again at the 105th second to the same location, the signal increased to 115 counts within 60 seconds, clearly demonstrating leak localization and dynamics.

\onecolumngrid

\twocolumngrid

\section{CONCLUSIONS}
\begin{figure}[!bh]
  \centering
  \includegraphics[width=0.48\textwidth]{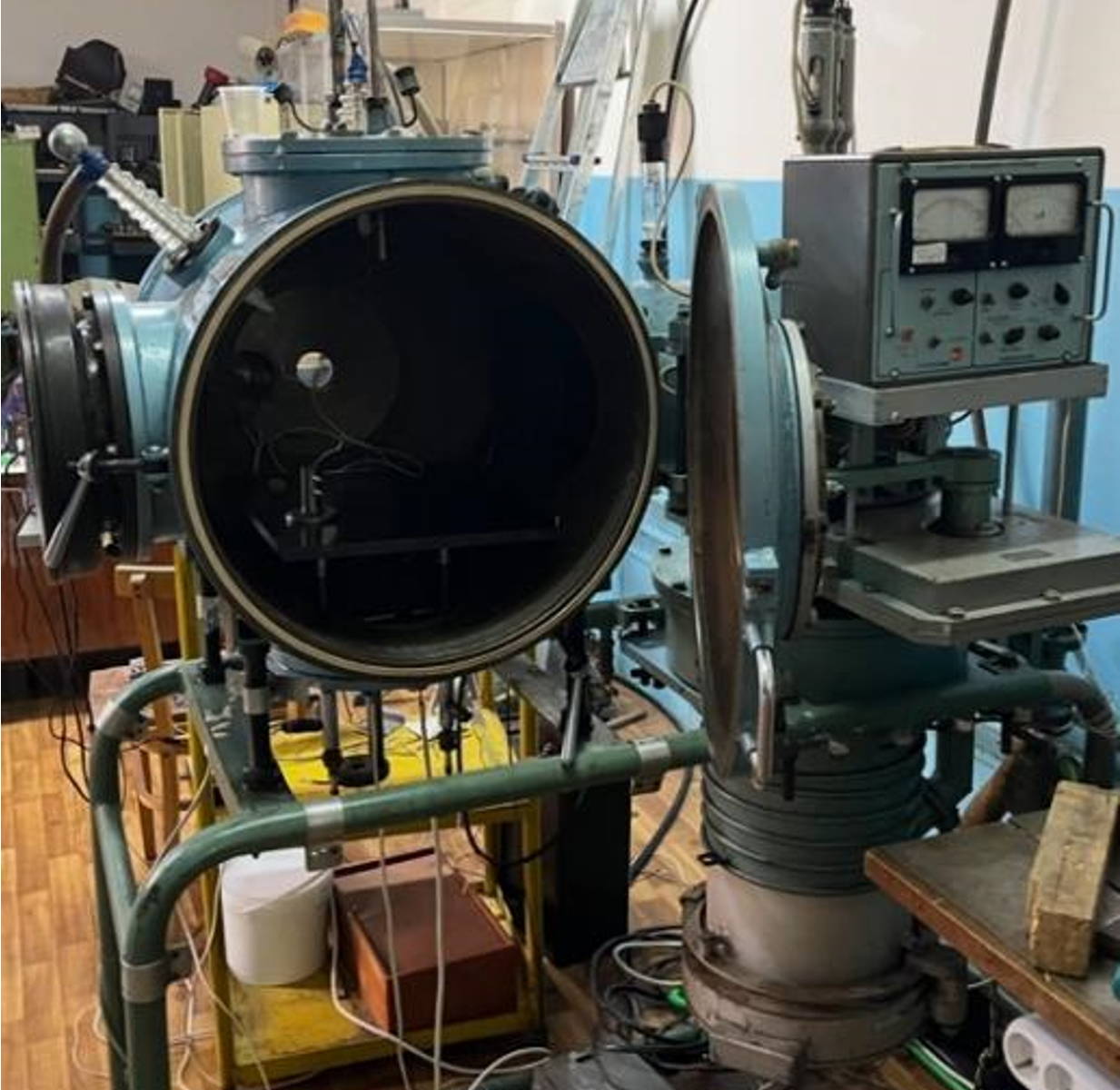}
  \caption{Vacuum setup "KASPAR".}
  \label{Fig:06}
\end{figure}

The developed device based on the MQ-8 sensor and Arduino platform demonstrated efficiency as a simple and low-resource tool for detecting leaks in vacuum systems. The tests were conducted on our vacuum setup "KASPAR" (\Figref{Fig:06}), representing a cylindrical chamber of 200 L volume equipped with a forevacuum pump 2NVR-5DM and a diffusion pump, with pressure measurement using thermocouple and ionization gauges VIT-2. The device proved convenient in operation, provided real-time data visualization, and allowed recording of hydrogen entering the working area of the setup. Initial measurements revealed pressure instability and the inability to achieve the target vacuum level of $10^{-4}$--$10^{-5}$ Pa, suggesting the presence of leaks. With the developed device, the leakage areas were localized and subsequently eliminated, after which the vacuum chamber was restored to a pressure of about $4\times10^{-5}$ Pa. After eliminating leaks, the pumping rate increased significantly, ensuring stable and rapid achievement of the required vacuum for further experiments on dielectric laser acceleration (DLA) of electrons. The device can also be used in other laboratory vacuum chambers due to its compactness, simplicity, low production cost, and convenience and safety of transport and operation, despite the use of hydrogen during leak detection.

The working range of the sensor was 100--10,000 ppm, which corresponds to the technical documentation data. During the experiments, the sensitivity to other gases (methane, butane) was also tested, confirming cross-selectivity. However, the highest response was observed specifically for hydrogen, which allows the developed leak detector to be considered as an affordable alternative to helium mass-spectrometric leak detectors in laboratory conditions, where simplicity and minimal cost are critical.

\begin{acknowledgments}
The study is supported by the National Research Foundation of Ukraine under the program "Excellent Science in Ukraine" (project \# 2023.03/0182).
\end{acknowledgments}

\end{document}